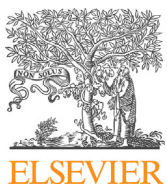
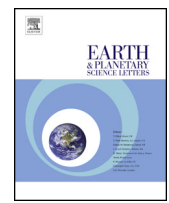

# Solubility of water in peridotite liquids and the prevalence of steam atmospheres on rocky planets

Paolo A. Sossi [a,*], Peter M.E. Tollan [a], James Badro [b], Dan J. Bower [c]

[a] *Institute of Geochemistry and Petrology, ETH Zürich, 8092 Zürich, Switzerland*
[b] *Institut de Physique du Globe de Paris, Université de Paris, 75005 Paris, France*
[c] *Center for Space and Habitability, Universität Bern, 3012 Bern, Switzerland*

**ABSTRACT**

Atmospheres are products of time-integrated mass exchange between the surface of a planet and its interior. On Earth and other planetary bodies, magma oceans likely marked significant atmosphere-forming events, during which both steam- and carbon-rich atmospheres may have been generated. However, the nature of Earth's early atmosphere, and those around other rocky planets, remains unclear for lack of constraints on the solubility of water in liquids of appropriate composition. Here we determine the solubility of water in 14 peridotite liquids, representative of Earth's mantle, synthesised in a laser-heated aerodynamic levitation furnace. We explore oxygen fugacities ($f\mathrm{O}_2$) between −1.9 and +6.0 log units relative to the iron-wüstite buffer at constant temperature (2173 ± 50 K) and total pressure (1 bar). The resulting $f\mathrm{H}_2\mathrm{O}$ ranged from nominally 0 to 0.027 bar and $f\mathrm{H}_2$ from 0 to 0.064 bar. Total $\mathrm{H}_2\mathrm{O}$ contents were determined by transmission FTIR spectroscopy of doubly-polished thick sections from the intensity of the absorption band at 3550 cm$^{-1}$ and applying the Beer-Lambert law. The mole fraction of water in the liquid is found to be proportional to $(f\mathrm{H}_2\mathrm{O})^{0.5}$, attesting to its dissolution as OH. The data are fitted by a solubility coefficient of 524 ± 16 ppmw/bar$^{0.5}$, given a molar absorption coefficient, $\varepsilon_{3550}$, of 6.3 ± 0.3 m$^2$/mol for basaltic glasses or 647 ± 25 ppmw/bar$^{0.5}$, for a preliminary $\varepsilon_{3550} = 5.1 \pm 0.3$ m$^2$/mol for peridotitic glasses. These solubility constants are roughly 10 - 25% lower than those for basaltic liquids at 1623 K and 1 bar. Higher temperatures result in lower water solubility in silicate melts, wholly offsetting the greater depolymerisation of peridotite melts that would otherwise increase $\mathrm{H}_2\mathrm{O}$ solubility relative to basaltic liquids at constant temperature. Because the solubility of water remains high relative to that of $\mathrm{CO}_2$, steam atmospheres are rare, although they may form under moderately oxidising conditions on telluric bodies, provided sufficiently high H/C ratios prevail.

## 1. Introduction

Owing to the heat generated by planetary impacts (Melosh, 1990) and short-lived radioactive isotopes (Bhatia, 2021), magma ocean-forming events were likely commonplace during the main planet-building stage (up to ∼100 Myr after CAI; Chao et al., 2021). Such events are significant because they likely result in the wholesale resetting of any pre-existing primary atmosphere captured from the solar nebula (Genda and Abe, 2005), and engender secondary atmosphere growth owing to the efficiency with which the molten planetary interior transports mass to the atmosphere. The Rayleigh number of a liquid terrestrial mantle, ∼10$^{30}$, yields mixing timescales of the order of days- to weeks (Solomatov, 2000), rapid with respect to atmospheric cooling timescales, meaning equilibrium between the interior and its atmosphere is readily attained. The corollary is that heat transfer is also efficient, meaning magma oceans are periodic, short-lived phenomena occurring over ∼10$^3$ to 10$^7$ yr (Bower et al., 2019; Lebrun et al., 2013).

The nature of an atmosphere generated during a magma ocean stage on a given planet is key to establishing its long-term stability to photochemical destruction, condensation or escape, and whether it is amenable to the development of life (Catling and Kasting, 2017). The identities and partial pressures of gas species that comprise an atmosphere depend upon the *i*) bulk planetary endowment of atmosphere-forming elements (typically H, C, N, O and S), and *ii*) the solubility of these gases in the underlying magma ocean as a function of the intensive variables (pressure, temperature, composition) at the interface. The bulk silicate Earth

\* Corresponding author.
*E-mail address:* paolo.sossi@erdw.ethz.ch (P.A. Sossi).

(BSE) acts as a reference model for the estimation of abundances of atmosphere-forming elements (*e.g.*, Hirschmann, 2018; Marty, 2012). Provided appropriate solubility laws are known, the composition of a model atmosphere in equilibrium with a magma ocean can be determined.

In this manner, Sossi et al. (2020) demonstrated that, at 2173 K, a CO-rich atmosphere would have formed around the Earth for oxygen fugacities ($fO_2$) near the iron-wüstite (IW) buffer. Such atmospheres would cool and partially condense to give rise to $CO_2$-$N_2$-rich atmospheres at ambient temperatures. They highlighted that the type of atmosphere depends on $fO_2$, which, at a given temperature and pressure, determines the ratios of $fCO/fCO_2$ and $fH_2/fH_2O$. Using a similar approach, Gaillard et al. (2022) highlighted that, should equilibrium between the magma ocean and atmosphere take place at very low $fO_2$ (ΔIW-3 or lower), then $H_2$-rich atmospheres would form, whereas $SO_2$-rich atmospheres could prevail under oxidising conditions (>ΔIW+2). However, the accuracy of these models is contingent upon the extrapolation of solubility laws of volatile species in molten peridotite from experimental data in molten basalt, and at much lower temperatures.

The determination of the solubilities of major volatiles in silicate melts has been largely confined to moderate- to high-pressures relevant to the Earth's crust and upper mantle (Dalou et al., 2019; Dixon et al., 1995; Hirschmann et al., 2012), with only a handful at 1 bar (Newcombe et al., 2017). However, atmospheric pressures above magma oceans on rocky planets are likely to be below several hundred bar (*e.g.*, Elkins-Tanton, 2008; Gaillard et al., 2022; Sossi et al., 2020). Moreover, the basaltic (and more evolved) compositions examined in these studies melt below ∼1773 K (Gaillard et al., 2001; Moore et al., 1998). By contrast, the Earth's mantle and those of other rocky planets are broadly peridotitic, not basaltic, in composition (Unterborn et al., 2020) with liquidus temperatures of ∼2023 K at 1 bar (Takahashi, 1986). The difficulty of achieving such temperatures experimentally, and in quenching peridotitic compositions into homogeneous glasses (*cf.* Green, 1975) has thus far prevented solubility measurements in magma ocean-relevant liquids.

Here we explore the occurrence of steam ($H_2O$-rich) atmospheres above magma oceans, a state often imposed in planetary cooling models (Abe and Matsui, 1985; Fegley et al., 2016; Lebrun et al., 2013). In order to do so, molten peridotite representative of Earth's mantle was equilibrated at 2173 K in a controlled atmosphere achieved by mixing Ar-$CO_2$-$H_2$ and $O_2$ gases in an aerodynamic laser-heated levitation furnace (ALLF) that permitted rapid quenching. The water contents and speciation in the resulting glasses were determined by Fourier-Transform InfraRed (FTIR) spectroscopy. Together, these data are used to devise a water solubility model for peridotite liquids and applied to understand atmospheres generated in equilibrium with magma oceans.

## 2. Methods

### 2.1. Experimental

The quenched peridotite liquids investigated in this study are those reported in Sossi et al. (2020). Briefly, a composition modelled on that of KLB-1 (Takahashi, 1986) was synthesised by mixing high-purity (>99.95% pure) powders of MgO, $SiO_2$, $CaCO_3$, $Fe_2O_3$ and $Al_2O_3$ together in the desired proportions. Following 24 h of decarbonation at 1273 K in air, ∼20 mg pieces of pressed powder were placed in the ALLF (Institut de Physique du Globe de Paris, France). All samples were heated above their liquidus by a 125 W $CO_2$ laser ($\lambda = 10.6$ μm) to 2173 ± 50 K for 30 s in pure $O_2$ before being quenched and re-melted under the desired gas mixture at the same temperature, as recorded by optical pyrometry. Gas flow rates were set to a precision of ±2% relative by Bronckhorst mass flow controllers delivering 92% Ar/8% $H_2$ and 92% Ar/8% $CO_2$ mixtures to the furnace nozzle, resulting in oxygen fugacities from ΔIW-1.9 to ΔIW+6.0. Gas flow rates were in the range 800 – 1000 sccm, resulting in nozzle velocities of 8 – 10 m/s (*cf.* Badro et al., 2021). Badro et al. (2021) and Young et al. (2022) showed that isotopic fractionation during evaporation in levitation furnaces results from the advective transport of the gas around the bead, resulting in partial pressures of evaporating species from the melt that are ∼75% lower than their equilibrium partial pressures. For a flowing gas representing an infinite reservoir to impose $fO_2$, $fH_2$ and $fH_2O$, the total pressure ($P_T$) around the sphere must be equal to 1 bar. At 2173 K, the temperatures are sufficiently high that the gas behaves ideally, such that the fugacity of species $i$, $f_i$, is equal to its partial pressure, $p_i$, calculated according to the equation $p_i = X_i P_T$, where $X$ is the mole fraction. The fugacities at the experimental temperature and pressure were calculated by Gibbs Free Energy minimisation using *FactSage*. Samples were quenched below the glass transition by switching off the power to the laser, resulting in cooling rates of ∼800 K/s. A summary of the experimental conditions can be found in Table 1. As reported by Sossi et al. (2020), the two most reduced samples experienced ∼10% Fe loss due to evaporation with a concomitant decrease in Si/Mg by ∼4% relative. The mean composition for the remaining peridotite glasses, by weight, is 46.53(26)% $SiO_2$, 4.37(4)% $Al_2O_3$, 8.44(29)% $FeO^T$ (all Fe as FeO), 38.05(42)% MgO, and 2.06(5)% CaO. Both $Na_2O$ and $K_2O$ are <0.01%.

### 2.2. Fourier-Transform InfraRed spectroscopy

Water contents of peridotite glasses were determined by FTIR spectroscopy at the Institut für Geologie, Universität Bern, Switzerland (Tollan and Hermann, 2019). The glasses were mounted in epoxy and cut to a total thickness of 1(±0.005) mm. The samples were then polished with diamond paste on both sides to a grade of 1 μm. Measurements were conducted in transmission mode under unpolarised light, using a Bruker Tensor II spectrometer with an SiC Globar infrared source and a KBr beamsplitter, coupled to a Bruker Hyperion 3000 microscope with a dry air-purged sample chamber. The infrared light transmitted through the sample was collected as a function of wavelength by a liquid $N_2$-cooled mercury cadmium telluride (MCT) LN-D316 detector, with an integration area of 400 × 15 μm per spot, 128 scans and 8 cm$^{-1}$ resolution over a spectral range from 6000 cm$^{-1}$ to 1000 cm$^{-1}$. For each sample, in-situ FTIR transects were performed across the diameter of the glasses, amounting to up to 10 spots/sample at a spacing of roughly 100 μm/point with the background determined every 20 points (Table S1, Fig. S1). The data were processed using the Bruker OPUS software, in which atmospheric absorbance from $H_2O$ and $CO_2$ was corrected for using its built-in algorithm. Owing to the low absorption intensity of the 3550 cm$^{-1}$ feature used for total water quantification and the steeply curved intrinsic absorption of the peridotitic glass, a polynomial baseline correction with ∼15 points was performed manually and independently three times, each ∼6 months apart. The contribution of the 3550 cm$^{-1}$ band to the absorbance, was quantified by its peak intensity relative to the local baseline within the assigned spectral window, which ranged from 3750 ± 50 cm$^{-1}$ to 3100 ± 50 cm$^{-1}$ for all spectra.

The Beer-Lambert law states that the total abundance of water dissolved in the glass, $C(H_2O)$, is related to the measured total absorbance of the IR-active band at 3550 cm$^{-1}$ ($I_{3550}$) produced by OH- and molecular $H_2O$ stretching modes (Stolper, 1982), its molar mass $M(H_2O)$ (in kg/mol), the path length of transmitted light through the absorbing medium ($d$ in m) and its density ($\rho$ in kg/m$^3$), by a proportionality constant, the molar absorption cross section for that band ($\varepsilon_{3550}$), with units of m$^2$/mol.

$$C(\text{H}_2\text{O})(ppm) = 10^6 \frac{I_{3550} M(\text{H}_2\text{O})}{d \rho \varepsilon_{3550}} \quad (1)$$

Here, $M(\text{H}_2\text{O}) = 0.018015$ kg/mol, $d = 10^{-3}(\pm 5 \times 10^{-6})$ m, and $\rho = 2991 \pm 11$ kg/m$^3$ at the glass transition temperature, $T_g = 1012 \pm 6$ K (Dingwell et al., 2004), calculated using thermal expansivities and partial molar volumes from (Lange and Carmichael, 1987) and compressibilities from (Kress and Carmichael, 1991), assuming Fe$^{3+}$/$\Sigma$Fe = 0 and anhydrous compositions for all samples. The calculated density extrapolated to 2173 K is in excellent agreement with synchrotron X-ray absorption measurements of peridotitic liquids, $\rho = 2700$ kg/m$^3$ (Sakamaki et al., 2010). The molar absorption cross section has been determined for the 3550 cm$^{-1}$ band in a wide range of silicate compositions, from basaltic to rhyolitic (Shishkina et al., 2014; Stolper, 1982) yet that for peridotitic compositions is unknown. Stolper (1982) derived a value of 6.71 $\pm$ 0.67 m$^2$/mol for $\varepsilon_{3550}$ for basaltic- to rhyolitic glasses, while Shishkina et al. (2014) determined $\varepsilon_{3550} = 5.92 \pm 0.40$ m$^2$/mol for the alkali basaltic and basanitic compositions, suggesting little dependence of $\varepsilon_{3550}$ on melt composition over this range. On the other hand, Mercier et al. (2010) concluded on the basis of SIMS and FTIR analyses of basanitic glasses that increasing non-bridging oxygen over tetrahedral oxygens (NBO/T) in the melt leads to a decrease in $\varepsilon_{3550}$ to roughly 4.5 m$^2$/mol at NBO/T $\sim$ 1. The water contents of the peridotite glasses were too low to permit accurate independent determination of $C(\text{H}_2\text{O})$ by Karl-Fischer Titration or by Elastic Recoil Detection Analysis (ERDA), which has a detection limit near $\sim$50 - 100 ppm (Bureau et al., 2009; Withers et al., 2012). However, thanks to the recent development of a novel rapid-quench multi-anvil technique, Bondar et al. (2021) were able to synthesise a set of hydrous peridotitic glasses with MgO contents up to 43.5 wt.% and water contents ranging from 0 to 5 wt.% H$_2$O (Bondar et al., 2022). On the basis of the hydrous peridotitic glass with the lowest water content, Bondar (2022) obtained a preliminary molar absorption coefficient for $\varepsilon_{3550}$ of 5.1 $\pm$ 0.3 m$^2$/mol. For completeness, we present H$_2$O concentrations calculated according to a mean value $\varepsilon_{3550} = 6.3 \pm 0.3$ m$^2$/mol ($n = 8$) for basaltic glasses, and the preliminary value of 5.1 $\pm$ 0.3 m$^2$/mol for peridotitic glasses. The uncertainties on these quantities are propagated to those on $C(\text{H}_2\text{O})$, reported as $1\sigma$.

## 3. Results

A summary of the results is shown in Table 1 and raw, unnormalised spectra for each sample are presented as Extended Data. Fig. 1 shows the calculated gas fugacities of the major species present in the gas mixes used in the experiments at a constant temperature of 2173 K. Owing to the binary mixture of CO$_2$-H$_2$ (diluted with 92% Ar), CO$_2$ is the dominant gas species for oxygen fugacities 1.5 log units above the IW buffer, while $f$H$_2$ predominates below IW. In the interval between IW and $\Delta$IW+1.5, H$_2$O is the prevailing gas species in the C-O-H system. The calculated $f$H$_2$O at 2173 K ranges from nominally 0 bar (Per-1, -2 and -5) to 0.027 bar (Per-10), while $f$H$_2$ extends up to 0.064 bar (Per-12) (Table 1).

The only H-bearing absorption band observed in the glasses is that located at 3550 cm$^{-1}$. Hence, we find no evidence for the dissolution of molecular H$_2$ at 4130 cm$^{-1}$. Within a sample, the normalised absorbance of the 3550 cm$^{-1}$ band varies by <7% relative, with an average of 2.8 $\pm$ 1.7%, indicating water is homogeneously distributed in the glasses (Table S1). Moreover, $I_{3550}$ scales with $f$H$_2$O, reaching a maximum at $\sim$0.5 log units above the iron-wüstite buffer (Fig. 2). The corresponding water contents of peridotite glasses as determined by eq. (1) range from 32.5 $\pm$ 2.3 ppmw for $\varepsilon_{3550} = 6.3$ m$^2$/mol or 40.2 $\pm$ 3.2 ppmw for $\varepsilon_{3550} = 5.1$ m$^2$/mol (Per-5) to 137.4 $\pm$ 7.3 ppmw ($\varepsilon_{3550} = 6.3$ m$^2$/mol) or 169.7 ± 11.1 ppmw ($\varepsilon_{3550} = 6.3$ m$^2$/mol) (Per-11). By examining the water contents determined amongst three glasses synthesised at the same experimental conditions (Per-TS1, Per-7, Per-TS2) but for different times (10 s, 30 s, and 120 s, respectively), equilibrium distribution of $H_2O$ between the liquid and the surrounding gas phase is achieved in under 10 s. The $H_2O$ content in each of the three samples overlaps within uncertainty, with a relative standard deviation of 5.6%, on par with that observed for a single sample across the three different fitting schemes. Based on these replicate experimental samples and baseline fitting procedures, the total precision on $C(H_2O)$ is estimated to be ±5% relative (Table 1).

**Table 1**
Experimental conditions, calculated gas fugacities, measured infrared absorbance intensity of the 3550 cm$^{-1}$ band and calculated H$_2$O contents.

| Sample | Temperature (K) | Time (s) | $\Delta$IW[a] | $f$H$_2$ (bar) | $f$H$_2$O (bar) | # of points | Fit 1 Absorbance | Fit 2 Absorbance | Fit 3 Absorbance | Mean $\varepsilon = 6.3$ m$^2$/mol (ppmw) | $1\sigma$ | Mean $\varepsilon = 5.1$ m$^2$/mol (ppmw) | $1\sigma$ | Corrected Mean $\varepsilon = 6.3$ m$^2$/mol (ppmw) | Corrected Mean $\varepsilon = 5.1$ m$^2$/mol (ppmw) |
|---|---|---|---|---|---|---|---|---|---|---|---|---|---|---|---|
| Per-1 | 2173 $\pm$ 21 | 30 | 5.97 | 0 | 0 | 3 | 0.036 | 0.037 | 0.038 | 35.4 | 2.0 | 43.7 | 3.0 | 2.9 | 3.5 |
| Per-2 | 2166 $\pm$ 40 | 40 | 3.46 | 0 | 0 | 7 | 0.039 | 0.040 | 0.039 | 37.6 | 2.0 | 46.5 | 3.1 | 5.1 | 6.3 |
| Per-3 | 2134 $\pm$ 29 | 32 | 3.42 | 1.28E-05 | 0.00071 | 7 | 0.053 | 0.054 | 0.050 | 50.0 | 3.2 | 61.8 | 4.6 | 17.5 | 21.7 |
| Per-4 | 2197 $\pm$ 59 | 60 | 3.46 | 9.88E-07 | 5.74E-05 | 7 | 0.041 | 0.040 | 0.039 | 38.2 | 2.2 | 47.2 | 3.2 | 5.7 | 7.1 |
| Per-5 | 2239 $\pm$ 25 | 36 | 3.46 | 5.17E-08 | 3.02E-06 | 6 | 0.036 | 0.033 | 0.033 | 32.5 | 2.3 | 40.2 | 3.2 | 0.0 | 0.0 |
| Per-6 | 2151 $\pm$ 23 | 25 | 3.37 | 2.77E-05 | 0.0015 | 8 | 0.053 | 0.055 | 0.057 | 52.6 | 3.2 | 65.0 | 4.7 | 20.1 | 24.8 |
| Per-7 | 2139 $\pm$ 13 | 27 | 3.23 | 8.41E-05 | 0.0038 | 7 | 0.069 | 0.070 | 0.072 | 67.2 | 3.8 | 83.1 | 5.6 | 34.7 | 42.9 |
| Per-8 | 2197 $\pm$ 13 | 31 | 2.94 | 0.00024 | 0.0076 | 8 | 0.088 | 0.090 | 0.094 | 86.7 | 5.2 | 107.1 | 7.6 | 54.2 | 66.9 |
| Per-9 | 2175 $\pm$ 15 | 30 | 2.03 | 0.0016 | 0.018 | 9 | 0.115 | 0.120 | 0.118 | 112.5 | 6.3 | 139.0 | 9.4 | 80.0 | 98.8 |
| Per-10 | 2173 $\pm$ 21 | 30 | 0.64 | 0.012 | 0.027 | 9 | 0.141 | 0.142 | 0.143 | 135.8 | 7.1 | 167.7 | 10.9 | 103.3 | 127.5 |
| Per-11 | 2169 $\pm$ 16 | 28 | $-$0.77 | 0.041 | 0.018 | 10 | 0.145 | 0.142 | 0.144 | 137.4 | 7.3 | 169.7 | 11.1 | 104.8 | 129.5 |
| Per-12 | 2112 $\pm$ 17 | 33 | $-$1.90 | 0.064 | 0.0078 | 10 | 0.133 | 0.132 | 0.138 | 128.4 | 7.3 | 158.6 | 10.8 | 95.9 | 118.5 |
| Per-TS1 | 2124 $\pm$ 13 | 10 | 3.23 | 8.41E-05 | 0.0038 | 8 | 0.066 | 0.066 | 0.068 | 63.7 | 3.5 | 78.7 | 5.2 | 31.2 | 38.6 |
| Per-TS2 | 2148 $\pm$ 41 | 120 | 3.23 | 8.41E-05 | 0.0038 | 8 | 0.064 | 0.074 | 0.071 | 66.6 | 5.6 | 82.3 | 7.6 | 34.1 | 42.1 |

[a] IW buffer calculated according to the revised expression of Hirschmann (2021) and hence the delta values shown here are 0.5 log units lower than reported in Sossi et al. (2020).

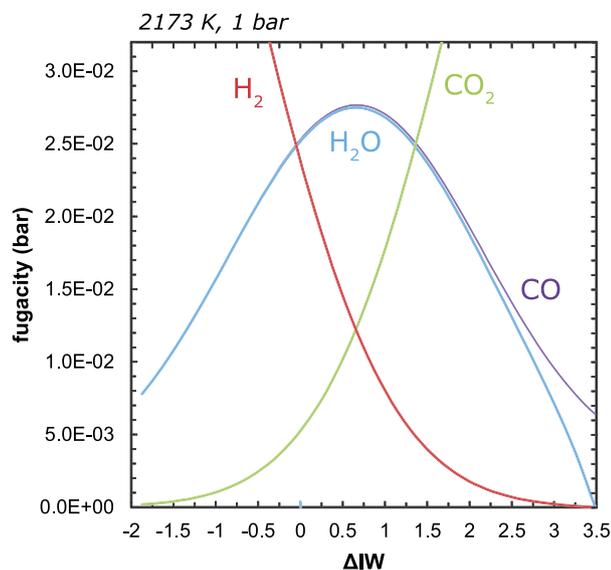

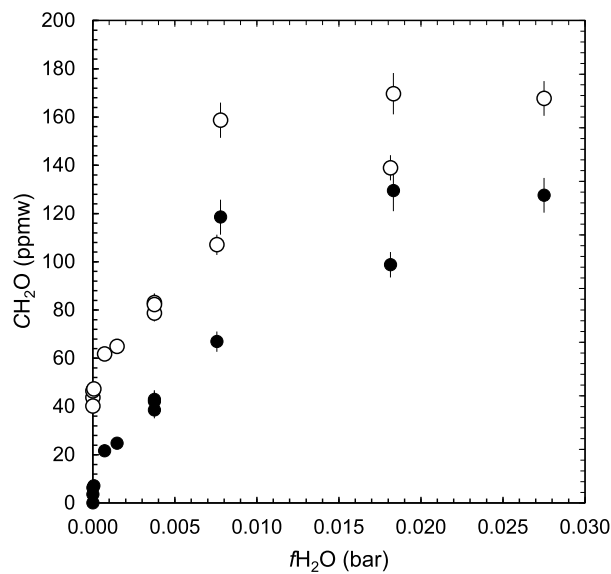

**Fig. 1.** Equilibrium speciation of the gas phase as calculated in *FactSage* for the $CO_2:H_2$ mixtures used in the experiments, from 0.11 to pure $CO_2$, in 92% Ar, at 2173 K and 1 bar, as a function of $fO_2$ relative to the IW buffer ($\Delta IW$) of Hirschmann (2021).

**Fig. 3.** Water contents of the glasses (in ppmw using $\varepsilon_{3550} = 5.1$ m$^2$/mol) expressed as a function of its fugacity in the gas phase at 2173 K. White points indicate uncorrected data, showing the water contents remain at ~40 ppmw for nominal $fH_2O = 0$, and black points show the data corrected according to eq. (2).

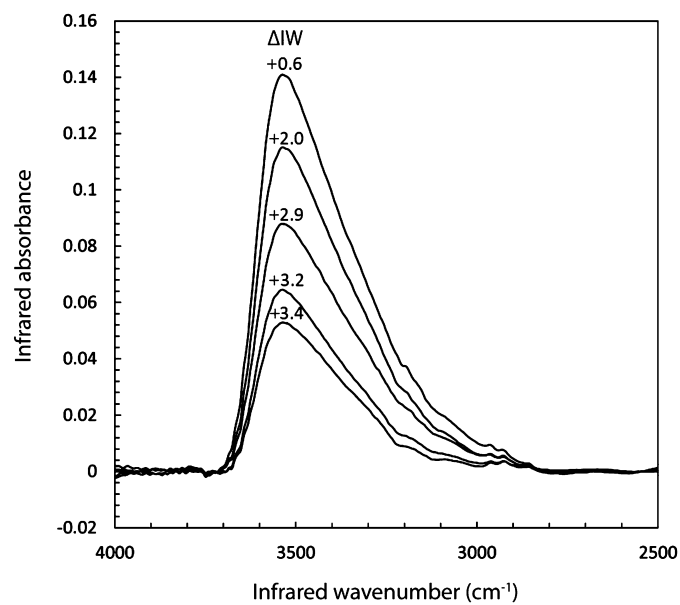

**Fig. 2.** Background-normalised FTIR spectra (see section 2.2) showing the intensity of the OH stretching absorbance band (~3550 cm$^{-1}$) for glasses over a range of $fO_2$, from $\Delta IW+0.6$ to +3.4 (where IW = oxygen fugacity defined by the iron-wüstite buffer at 2173 K).

When $C(H_2O)$ is plotted against $fH_2O$ (Fig. 3), it is observed that dissolved water contents do not tend to nil at $fH_2O = 0$ as expected thermodynamically, but rather reach a constant value of ~32 - 40 ppmw. As the $C(H_2O)$ must formally be equal to zero when $fH_2O = 0$, the $C(H_2O)$ contents of all samples are re-normalised to those of samples with $fH_2O = 0$;

$$C(H_2O)_{norm} = C(H_2O)_{meas} - C(H_2O)_0 \qquad (2)$$

Here we adopt the lowest $C(H_2O)$ value measured in the glasses, 32.5 ppmw ($\varepsilon_{3550} = 6.3$ m$^2$/mol) or 40.2 ppmw ($\varepsilon_{3550} = 5.1$ m$^2$/mol), as the normalising value, $C(H_2O)_0$. Alternatively, the non-negligible absorbance for samples with nominal $fH_2O$ of zero could result from a constant $fH_2O$ background during glass synthesis. However, when excluding these samples on a plot of $\log fH_2O$ vs. $\log CH_2O$, the data define a slope of ~0.33, well below the value of 0.5 expected for dissolution of OH in the melt. This explanation is therefore untenable, as it implies a gas-liquid reaction that does not relate to any physical dissolution stoichiometry. Moreover, $H_2O$ contents of the ALPHAGAZ $O_2$ and Ar-$CO_2$ mixtures used in the experiments were <3 ppmw, which would translate into ~1 ppmw dissolved in the melt using pre-existing calibrations (Newcombe et al., 2017). As the experiments were not performed within a chamber, it cannot be excluded that the melt bead interacted with small quantities of air during the experiment, as levitation sometimes becomes unstable during laser power and/or gas fluctuations. Therefore, the remnant absorption at 3550 cm$^{-1}$ in melts synthesised under $fH_2O = 0$ is attributed to background contributions from either the environment or non-equilibrium water in the glass. That this water likely resides in the glass is supported by the fact that the measurements were performed in transmission mode, traversing the sample, and that the spectral peak shape and position are similar to those glasses that contain higher $H_2O$ contents. It is noteworthy that Newcombe et al. (2019) also observed 6 – 22 ppmw $H_2O$ dissolved in glasses equilibrated in nominally dry CO-$CO_2$ gas mixtures, which they attributed to kinetic inhibition of $H_2O$ degassing. The corrected values in Table 1 will be used for the remainder of the study.

The stoichiometry of the dissolution of water in silicate melts can be inferred from the dependence of $CH_2O$ on $fH_2O$. The dependence of water content in a liquid in equilibrium with a gas phase can be described:

$$C(H_2O) = \alpha [f(H_2O)]^\beta \qquad (3)$$

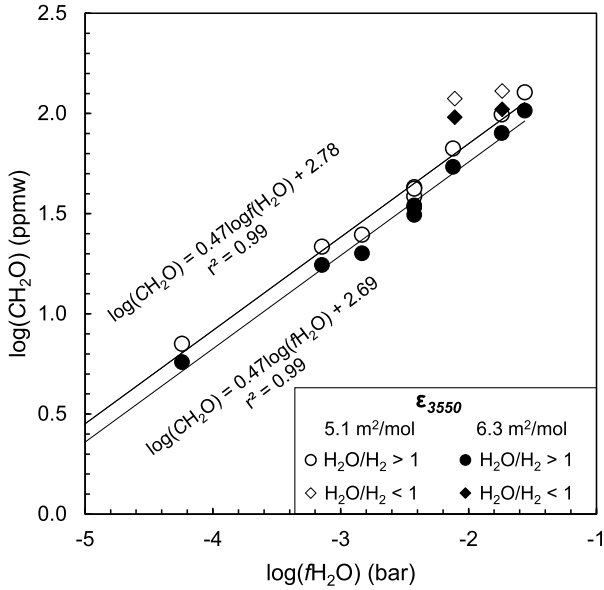

**Fig. 4.** Fit of equation (4) to the water contents (in ppmw) determined in the glass and the corresponding water fugacities in the gas phase. Water contents are calculated with $\varepsilon_{3550} = 5.1$ m$^2$/mol (white) and with $\varepsilon_{3550} = 6.3$ m$^2$/mol (black). Points shown in diamonds correspond to samples synthesised under high $fH_2$ and were not included in the regression. Samples that were synthesised at $fH_2O = 0$ (Per-1, Per-2, Per-5) were excluded from the plot.

Where $\alpha$ is the solubility coefficient and $\beta$ the stoichiometric coefficient of the reaction; Henry's Law is recovered for $\beta = 1$. The values of constants $\alpha$ and $\beta$ are determined empirically by taking the logarithm of both variables (Fig. 4);

$$\log C(H_2O) = \beta \log[f(H_2O)] + \log \alpha \qquad (4)$$

A least-squares regression of the data ($n = 9$) to eq. (4) yields $\beta = 0.47 \pm 0.02$ and, for a peridotite glass molar absorption coefficient ($\varepsilon_{3550} = 5.1$ m$^2$/mol), $\log \alpha$ of $2.78 \pm 0.05$ ($\alpha = 605^{+68}_{-60}$ ppmw/bar$^{0.47}$) while for $\varepsilon_{3550} = 6.3$ m$^2$/mol, $\log \alpha = 2.69 \pm 0.07$ ($\alpha = 490^{+68}_{-60}$ ppmw/bar$^{0.47}$) (Fig. 4). Two samples synthesised at the lowest $fO_2$ and highest $fH_2$ (Per-11 and Per-12) plot to markedly higher $CH_2O$ than those predicted by eq. (4) (see Fig. 4) and were excluded from the fitting procedure, along with samples synthesised at $fH_2O = 0$ (Per-1, Per-2 and Per-5). The accuracy of equation (4) is contingent on the assumption that the quantity of H in the glasses depends exclusively on $fH_2O$; one that will be discussed below.

## 4. Discussion

### 4.1. Mechanisms and equilibria of H dissolution

The value of $\beta$ obtained from the fit to the experimental data is approximately 0.5, a solubility dependence known as Sieverts' Law (Sieverts, 1929). This implies dissociation of the dissolved species in the liquid phase, a mechanism that was well-established in earlier experiments of water solubility in silicate melts in equilibrium with a gas or fluid phase (Dixon et al., 1995; Shishkina et al., 2010; Stolper, 1982; Tomlinson, 1956). More recently, similar results were obtained by Newcombe et al. (2017), who also observed a dependence of $H_2O$ content of lunar- and anorthite-diopside liquids equilibrated in a 1 bar furnace with $CO_2$-$H_2$ gas mixtures on $f(H_2O)^{0.5}$. Taken together, these solubility relationships are strong evidence for the dissolution of water into silicate melts by

$$H_2O(g) = H_2O(l) \qquad (5)$$

and the homogeneous equilibrium in the melt for its dissociation into hydroxyl groups

$$H_2O(l) + O(l) = 2OH(l) \qquad (6)$$

Where, at equilibrium, eq. (5) and (6) combine to yield:

$$K_{(7)} = \frac{a(OH)^2}{f(H_2O)a(O)}. \qquad (7)$$

Here, $a$ refers to the activity and states that the mole fraction of OH dissolved in the melt should be proportional to $(fH_2O)^{0.5}$. However, as shown in Fig. 4, the $H_2O$ contents of samples Per-11 and Per-12 do not abide by this theoretical expectation. This mismatch implies the contribution of a second equilibrium;

$$H_2(g) = H_2(l) \qquad (8)$$

because $H_2$ is the predominant H-bearing species in the gas phase in equilibrium with Per-11 and Per-12 (Table 1). Given that OH is expected to be the prevailing species at low $C(H_2O)$ (Newcombe et al., 2017) and no molecular hydrogen absorption band was observed in the spectra, a dissociation reaction equivalent to that of eq. (6) can be written for $H_2$,

$$H_2(l) + 2O(l) = 2OH(l) \qquad (9)$$

leading to:

$$K_{(10)} = \frac{a(OH)^2}{f(H_2)a(O)^2} \qquad (10)$$

Reaction (10) illustrates how the presence of $H_2(g)$ contributes to the dissolution of H as OH in peridotite liquids. The dataset (excluding samples Per-1, -2 and -5) is therefore fit by least squares regression to equation (11) that accounts for the contributions of both $fH_2$ and $fH_2O$ in the atmosphere, given a $\beta_{H_2O} = \beta_{H_2} = 0.5$ dependence (eq. (7), (10)) and varying $\alpha_{H_2O}$ and $\alpha_{H_2}$. This yields an expression for the solubility of $H_2O$ in peridotite liquids at 2173 K and 1 bar total pressure:

$$C(H_2O)(ppm) = (\alpha_{H_2O})f(H_2O)^{0.5} + (\alpha_{H_2})f(H_2)^{0.5} \qquad (11)$$

For $\varepsilon_{3550} = 6.3$ m$^2$/mol, we find $\alpha_{H_2O} = 524 \pm 16$ ppmw/bar$^{0.5}$ and $\alpha_{H_2} = 183 \pm 14$ ppmw/bar$^{0.5}$ whereas with $\varepsilon_{3550} = 5.1$ m$^2$/mol, these values are $\alpha_{H_2O} = 647 \pm 25$ ppmw/bar$^{0.5}$ and $\alpha_{H_2} = 225 \pm 13$ ppmw/bar$^{0.5}$. The goodness of fit with respect to the data is shown in Fig. 5. The solubility constant for $H_2O$, is roughly 10% (5.1 m$^2$/mol) to 25% (6.3 m$^2$/mol) lower than that observed for a lunar green glass (LG) and anorthite-diopside eutectic (ADeu) melts equilibrated at 1623 K (Newcombe et al., 2017). Unlike Newcombe et al. (2017), who are unable to detect any increase in OH solubility in the presence of $H_2(g)$ in LG and ADeu melts, peridotite liquids in equilibrium with $H_2(g)$ are able to dissolve OH with a solubility constant a factor of ∼3 lower than that for $H_2O(g)$.

Comparison with models based on pre-existing data for $H_2O$ solubility should only be considered qualitative because the experiments fall well outside their temperature calibration range. Nevertheless, for completeness, the model of Moore et al. (1998) (973 < T [K] < 1473) underpredicts the $H_2O$ contents by a factor ∼3, while that of Iacono-Marziano et al. (2012) (1373 < T [K] < 1673) overpredicts them by approximately the same factor (Table S2). Both serve to show that current models are inadequate for extrapolation beyond their calibration ranges, and applicability of such models to magma ocean-like conditions and compositions is not recommended.

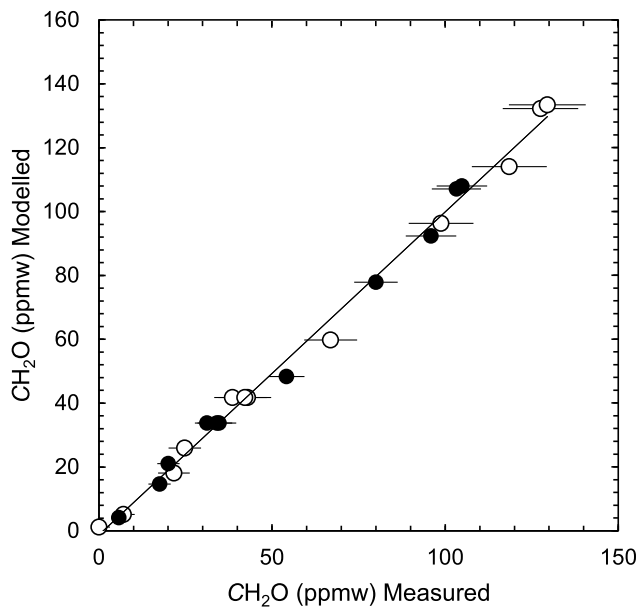

**Fig. 5.** Comparison between the measured water content of silicate melts determined in this work ($H_2O$, Measured) and the modelled value ($H_2O$, Modelled) determined using eq. (11) for $\varepsilon_{3550} = 5.1$ m$^2$/mol (white) and $\varepsilon_{3550} = 6.3$ m$^2$/mol (black). The fits are $H_2O$ (ppmw, modelled) = 1.01 × $H_2O$ (ppmw, measured) − 1.25, $r^2 = 0.995$ and $H_2O$ (ppmw, modelled) = 1.02 × $H_2O$ (ppmw, measured) − 1.6, $r^2 = 0.994$, respectively.

### 4.2. Thermodynamic constraints on the solubility of water at low $fH_2O$

Quantifying the influence of temperature and composition on the solubility of water in silicate melts requires evaluation of eq. (7). The lack of thermodynamic data for individual melt species restricts estimates of the equilibrium constant, $K_{(7)}$, to semi-empirical means (Hirschmann et al., 2012). The quantity of OH dissolved is given by:

$$X(OH) = \frac{\sqrt{K_{(7)} f(H_2O) a(O)}}{\gamma(OH)} \quad (12)$$

Where $X$ refers to the mole fraction and $\gamma$ to the activity coefficient. The activity coefficient of OH in silicate melts is not known, and is therefore assumed to be unity. Iacono-Marziano et al. (2012) considered that the activity of oxygen in silicate melts, O, is related to the ratio of Non-Bridging Oxygens (NBO) to total oxygen ($\Sigma$O). Here we define:

$$a(O) = \frac{NBO}{\Sigma O}, \quad (13)$$

where NBO and O are calculated following Iacono-Marziano et al. (2012). Given that $fH_2O = (X(OH)/\alpha_{XH_2O})^2$, (*cf*. eq. (3), where $X$ denotes $\alpha$ calculated by mole fraction of $H_2O$ in the melt) and $a = X\gamma$, one can rewrite eq. (7) as:

$$K_{(7)} = \frac{(\gamma(OH)\alpha_{XH_2O})^2}{a(O)} \quad (14)$$

This analysis illustrates that, all else being equal, $\alpha$ is expected to correlate positively with $a$O for a given pressure and temperature. That is, liquids with a greater proportion of Non-Bridging Oxygens should be able to dissolve more water at a given $fH_2O$, temperature and pressure. The peridotite liquids investigated here have $a$O = 0.43, while the LG and ADeu compositions of Newcombe et al. (2017) have lower NBO/$\Sigma$O, 0.35 and 0.28, respectively, reflecting their more polymerised melt structure. Because $\alpha$ is expected to increase with $(aO)^{0.5}$, which is opposite to observations, we conclude that composition plays only a subordinate role in determining the solubility constant of eq. (7) (*cf*. Shishkina et al., 2010; Iacono-Marziano et al., 2012), though systematic experimentation is required to further investigate compositional dependence at 1 bar total pressure.

By the same logic, to understand the effect of melt composition on the dissolution of OH derived from $H_2$, a reaction equivalent to eq. (14) may be written using $K_{(10)}$:

$$K_{(10)} = \frac{(\gamma(OH)\alpha_{XH_2})^2}{a(O)^2}. \quad (15)$$

Equation (15) illustrates that the solubility of OH in melts in equilibrium with $H_2(g)$ should increase proportional to $a$O. As a result, relative to the LG and ADeu compositions (Newcombe et al., 2017) with $a$O of 0.35 and 0.28, respectively, a 25 – 50% increase in OH solubility in peridotite liquid, with $a$O of 0.43, is expected. This increase is qualitatively compatible with the observation of a contribution of OH dissolution from the $H_2$ gas in peridotite liquids (*cf*. Fig. 4, eq. (11)) but not in the experiments of Newcombe et al. (2017). It is therefore likely that highly depolymerised melt compositions could incorporate large proportions of OH in their structure, even in the absence of $H_2O$ in the gas phase.

The other key variable, temperature, is poorly constrained because there are very few studies that report results of a wide experimental temperature series for a given composition at fixed $fH_2O$ or total pressure. Here we evaluate the dependence of water solubility on temperature at low total pressures and low $fH_2O$ such that the only dissolved species is OH (i.e., where eq. (7) applies). In evaluating $K_{(7)}$ from eq. (14), the value of $\alpha_{H_2O}$ is calculated on a molar basis, where $\alpha_{X(H_2O)} = \alpha_{H_2O} \times (5.55 \times 10^{-6})$, $\gamma$(OH) is assumed to be unity, and $a$O is calculated on the basis of melt composition (eq. (13)). The values of $K_{(7)}$ are calculated in this manner for five silicate melt compositions (Table 2). These values are converted to a Gibbs Free Energy change of reaction (7) by:

$$\Delta G_{(7)} = -RT\ln K_{(7)} \quad (16)$$

The resultant $\Delta G_{(7)}$ is plotted as a function of temperature in Fig. 6.

The enthalpy and entropy changes associated with the reaction are given by the intercept and slope, respectively, of Fig. 6. This is the 2$^{nd}$ Law Method, which implicitly assumes that heat capacity changes over the temperature range are sufficiently small so as to be considered constant. Thus, the values of $\Delta H = -54552 \pm 12500$ J/mol ($-32160 \pm 12500$ J/mol for $\varepsilon_{3550} = 5.1$ m$^2$/mol) and $\Delta S = 115.2 \pm 6.9$ J/mol.K (101.4 ± 6.9 J/mol.K for $\varepsilon_{3550} = 5.1$ m$^2$/mol) correspond to the mean temperature, 1800 K, over which the curve is fit. That the uncertainties on both quantities are relatively large reflects the three data points at 1 bar total pressure used to fit the relationship. However, using all data in Table 2, overlapping estimates are obtained, whereby $\Delta H = -61618 \pm 8044$ J/mol and $\Delta S = 116.1 \pm 5.6$ J/mol.K ($\Delta H = -53867 \pm 7540$ J/mol and $\Delta S = 110.3 \pm 5.3$ J/mol.K for $\varepsilon_{3550} = 5.1$ m$^2$/mol).

When translated to $H_2O$ dissolved in the silicate melt, a change in temperature from 1200 K to 2200 K results in a ∼2.5-fold decrease in water content of the melt. Notably, the model of Moore et al. (1998) predicts a similar temperature dependence (a factor ∼1.7 decrease over the same range), while the temperature term in the expression of Iacono-Marziano et al. (2012) is so small so as to be insignificant. The amount of $H_2O$ dissolved flattens out at high $fH_2O$, the implications of which are examined in the following sections.

**Table 2**
Compilation of experiments performed assessing water solubility in silicate melts at low total pressures over a range of temperatures and compositions. NB. The value of $\alpha_{H_2O}$ is calculated on the basis of mole fraction rather than ppmw.

| Reference | Composition | Temperature (K) | $\alpha$ (bar$^{-0.5}$) | $fH_2O$ (bar) | # of experiments |
|---|---|---|---|---|---|
| This work ($\varepsilon_{3550} = 6.3$ m$^2$/mol) | Peridotite | 2173 | $2.91 \times 10^{-3}$ | $5.7 \times 10^{-5}$ – 0.027 | 14 |
| This work ($\varepsilon_{3550} = 5.1$ m$^2$/mol) | Peridotite | 2173 | $3.59 \times 10^{-3}$ | $5.7 \times 10^{-5}$ – 0.027 | 14 |
| Newcombe et al. (2017) | Anorthite-Diopside eutectic | 1623 | $4.22 \times 10^{-3}$ | $9.8 \times 10^{-3}$ – 0.32 | 14 |
| Newcombe et al. (2017) | Lunar Green Glass | 1623 | $4.04 \times 10^{-3}$ | $9.8 \times 10^{-3}$ – 0.32 | 11 |
| Dixon et al. (1995) | Mid-Ocean Ridge Basalt | 1473 | $5.36 \times 10^{-3}$ | 17 – 709 | 14 |
| Hamilton and Oxtoby (1986) | NaAlSi$_3$O$_8$ | 1123 – 1573 | $7.59 \times 10^{-3}$ – $9.91 \times 10^{-3}$ | 1685 – 2160 | 13 |

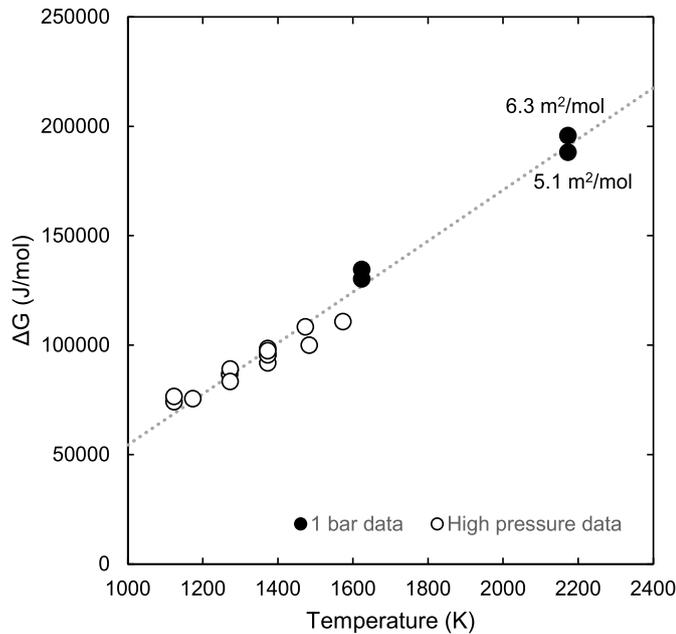

**Fig. 6.** The Gibbs Free Energy change of reaction (7) plotted as a function of temperature. Black = 1 bar data from this study and Newcombe et al. (2017); White = higher pressure data from Dixon et al. (1995) and Hamilton and Oxtoby (1986). These latter authors only provide total pressure, therefore, assuming a pure H$_2$O fluid in their experiments, total pressure was converted to $fH_2O$ using an equation of state (Duan and Zhang, 2006). Grey dotted line = best fit to all data.

## 5. Implications

### 5.1. Effect of planetary mass on the extent of water retention

The formation of planetary bodies in the early Solar System proceeds by accretion of material through impacts, which become progressively more energetic as the masses of the bodies involved increases. Moreover, as these processes occurred within the first several million years following the collapse of the molecular cloud that led to the protoplanetary disk, short-lived radioactive isotopes, most notably, $^{26}$Al with a half-life of 0.7 Myr, were still active (e.g., Bhatia, 2021). These provide sources of heat that can lead to extensive melting of the precursor rocky bodies of the Earth and the other terrestrial planets (Chao et al., 2021).

Records of these bodies are preserved in achondritic meteorites, such as eucrites from the asteroid IV-Vesta. These samples, which are broadly basaltic in composition, contain nominally anhydrous minerals, namely clinopyroxenes, that preserve low H$_2$O contents (∼5 – 10 ppmw; Sarafian et al., 2017, 2019; Stephant et al., 2021) with respect to their terrestrial counterparts (50 – 700 ppmw; Warren and Hauri, 2014). The uniformly low H$_2$O contents (<20 ppmw) are a general feature of clinopyroxenes from achondritic mafic meteorites from small telluric bodies, including angrites and ureilites (Peterson et al., 2022; Sarafian et al., 2017).

This observation, together with their negative- to chondritic $\delta$D ratios (McCubbin and Barnes, 2019), suggests that i) H originates from the accretion of the parent body and ii) the H preserved is not residual from partial evaporative loss, which would produce positive $\delta$D with respect to chondrites. Here we examine these constraints in the context of magma ocean outgassing on planetary bodies.

The partial pressure, $p$, exerted by a given gas species in a single-species atmosphere (see Bower et al., 2019 for a multi-species atmosphere), $i$, at a planetary surface is given by:

$$p_i = \frac{gM_i^a}{4\pi r^2}, \qquad (17)$$

where $M$ is the mass of $i$ in the atmosphere, $a$, and $r$ is the radius of the body. We model scenarios as a function of planetary radius in which the bulk water content is fixed (1000 ppmw), and can reside either in the magma ocean (assumed equivalent to the 70% the total mass of the body, i.e., comprised of a silicate mantle with a 30% core by mass) or in the atmosphere. This is expressed in the mass balance constraint:

$$M_i^T = M_i^a + M_i^m \qquad (18a)$$

where $T$ and $m$ refer to the total mass and mantle mass of $i$, respectively. Substituting equation (17) into equation (18a), and given that $M_i^T = M_T C_{i,T}$, where $C_{i,T}$ is the concentration (by mass) of $i$, one obtains:

$$M_T C_{i,T} = p_i \frac{4\pi r^2}{g} + M_m C_{i,m} \qquad (18b)$$

The bulk mass of the body, $M_T$, is calculated by solving the static structure equations that also yield gravity and internal pressure as a function of radius, $r$ (Valencia et al., 2007; Appendix A). The density of the mantle with depth is calculated assuming compression under adiabatic conditions, according to the Adams-Williamson equation, which accounts for the compressibility of typical mantle peridotite (Jackson, 1998). Finally, the quantity $C_{i,m}$ is eliminated by substitution of eq. (3), connecting the mantle abundance of H$_2$O to its partial pressure (note ideality in the gas phase is assumed, and $p$ is interchangeable with $f$, the fugacity):

$$M_T C_{H_2O,T} = fH_2O \frac{4\pi r^2}{g} + M_m \alpha_{H_2O} (fH_2O)^{0.5} \qquad (19)$$

Equation (19) is solved numerically, by varying $fH_2O$ such that the right-hand side equals the left-hand side.

Solutions to eq. (19) are represented graphically in Fig. 7. For a constant H$_2$O content of 1000 ppmw, partial pressures increase markedly above bodies with surface gravity up to ∼2 m/s$^2$, corresponding to radii of ∼2000 km. Over the same interval, the fraction of water outgassed decreases drastically, until the partial

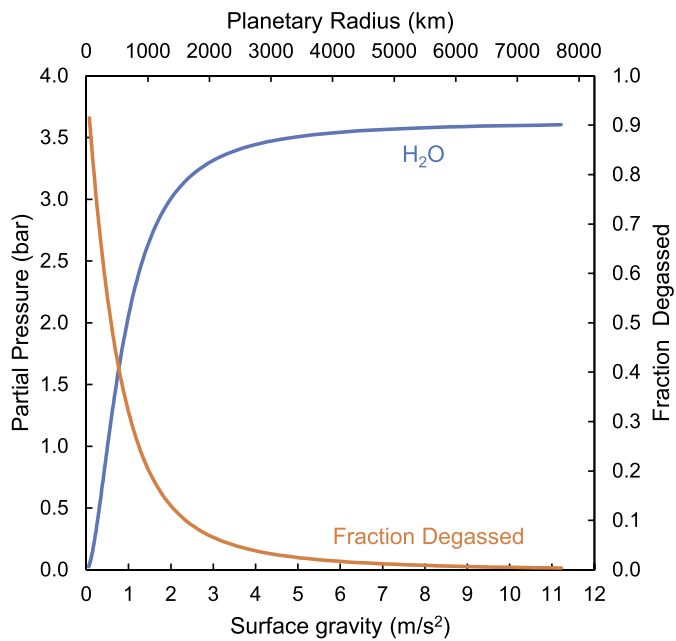

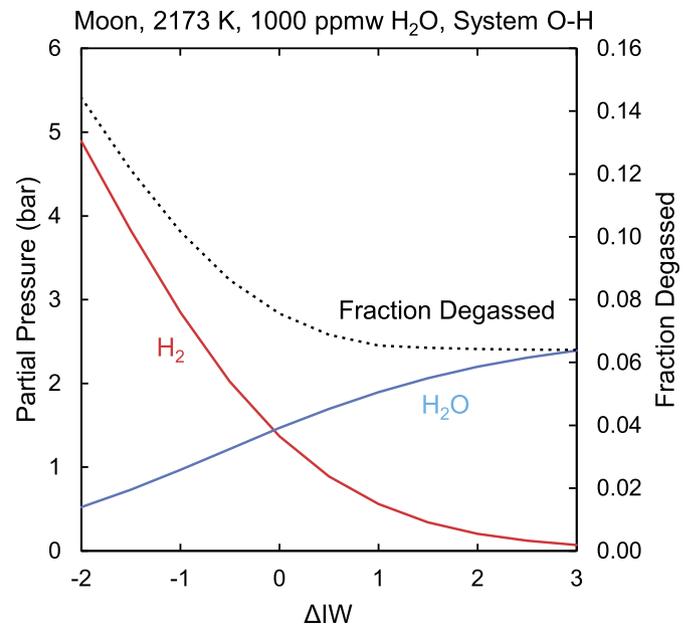

Fig. 7. The partial pressure of $H_2O$ (blue line) in a single-species atmosphere and the corresponding fraction of total water in the atmosphere (orange line) above a planetary surface as a function of its surface gravity and radius. While the absolute partial pressure increases with increasing planetary mass, the relative fraction of water present in the atmosphere (as opposed to dissolved in the liquid mantle) decreases. The calculations assume a fixed total water concentration (1000 ppmw as $H_2O$) for all planetary masses, a bulk density calculated by static structure equations assuming adiabatic compression (Appendix A), an $\alpha_{H_2O}$ of 524 ppmw/bar$^{0.5}$ (use of $\alpha_{H_2O} = 647$ ppmw/bar$^{0.5}$ would result in even lower pressures), a temperature of 2173 K and that the mass of the mantle is equal to 0.7 times the mass of the planet (i.e., an Earth-like core mass fraction). (For interpretation of the colours in the figure(s), the reader is referred to the web version of this article.)

Fig. 8. Partial pressures of H-bearing species in the system O-H in an hypothetical primitive lunar atmosphere at 2173 K as a function of oxygen fugacity expressed relative to the iron-wüstite buffer ($\Delta$IW). Red = $H_2$, Blue = $H_2O$. The initial concentration of $H_2O$ in the Moon is assumed to be that of Earth's mantle (1000 ppmw = $7.33 \times 10^{19}$ kg), its radius 1737 km and surface gravity = 1.62 m/s$^2$. An $\alpha_{H_2O}$ of 524 ppmw/bar$^{0.5}$ is used (use of $\alpha_{H_2O} = 647$ ppmw/bar$^{0.5}$ would result in even lower pressures).

pressure reaches an asymptote at roughly 3.5 bar for more massive planets.

As a result, the $H_2O$ budget can be lost more readily from smaller planetary bodies with respect to larger ones, despite lower partial pressures in their atmospheres, because a greater fraction of the water resides in the atmosphere. For example, magma oceans on bodies with the surface gravity of Vesta (0.25 m/s$^2$) would be expected to have ∼50% of their total $H_2O$ present in the atmosphere, compared with 0.5% for a body with the mass of Earth (Fig. 7). All else being equal, smaller bodies will tend to outgas atmospheres whose composition closely reflects that of their bulk planetary inventory, while species in the atmospheres of larger bodies will be moderated by their solubilities.

The foregoing analysis presumes all H is present as $H_2O$, and that the $fO_2$ is independent of planetary size. It is likely, however, that a significant fraction (∼50%) of H existed as $H_2$ in the atmospheres of small telluric bodies, given that the $fO_2$ recorded in basaltic achondrites is near the IW buffer (Wadhwa, 2008). Accordingly, because $H_2$ is less soluble than is $H_2O$ in silicate liquid (Hirschmann et al., 2012), the lower oxygen fugacities of small rocky bodies relative to Earth's present-day upper mantle ($\Delta$IW+3.5) make H loss from the former more pronounced. Therefore, the escape of $H_2O$ from small planetary bodies is not only favoured by their lower escape velocities, but also by their lower $fO_2$ and the solubility behaviour of $H_2O$ that results in proportionally greater degassing extents. These controls provide a physical explanation for the observation that all achondrites from small telluric bodies (Sarafian et al., 2019; Peterson et al., 2022) are strongly depleted in water with respect to the Earth in a manner inversely proportional to their radii (McCubbin and Barnes, 2019).

### 5.2. Water in the primitive atmosphere of the Moon

While similarities in the chemical and isotopic composition of refractory and major rock-forming elements in the Earth and Moon, such as Cr, Ti and O, have been used to argue for formation of the Moon directly from, or by complete mixing with Earth-like material (Lock et al., 2020; Pahlevan and Stevenson, 2007), this concordance breaks down for elements more volatile than Li, which are systematically depleted in the Moon (Tartèse et al., 2021). This depletion extends to hydrogen, whose abundance in the lunar mantle has been estimated through measurement of pyroclastic glasses (Saal et al., 2008) as well as from melt inclusions in mare basalts (Chen et al., 2015; Ni et al., 2019), to fall in the range 3 – 84 ppmw $H_2O$ (McCubbin and Barnes, 2019 and references therein). Though the range is wide, there is little doubt that the Moon is depleted in H with respect to Earth's mantle by a factor of 10 – 1000.

Nakajima and Stevenson (2018) argued that, in an $H_2O$-dominated protolunar atmosphere, H loss would be limited by its sluggish diffusion through the atmosphere with a mean molecular mass greater than that of $H_2$. However, these authors considered only $H_2O$-rich atmospheres ($fH_2O/fH_2 > 100$) which implicitly assumes oxidising conditions (at 2000 K, the $fH_2O/fH_2$ ratio reaches ∼100 at the Fayalite-Magnetite-Quartz, FMQ buffer). Depending on the temperature and $fO_2$ at which the atmosphere equilibrated with the magma, $fH_2/fH_2O$ could vary by orders of magnitude, potentially aiding the escape of H and other gases in the atmosphere.

Beginning with the premise that the Moon was derived from the Earth's mantle (Ringwood and Kesson, 1977), the initial concentration of $H_2O$ in the lunar mantle is expected to have been that of the present-day Earth's mantle, that is, roughly 1000 ppmw (Marty, 2012; Hirschmann, 2018). The formation of a primitive lunar atmosphere in equilibrium with a magma ocean at 2173 K in the simple system O-H as a function of $fO_2$ (Fig. 8) is simulated using eq. (11) for $H_2O$ and the data of Hirschmann et al. (2012)

for the dissolution of $H_2$. For this purpose, the homogeneous gas phase equilibrium $H_2 + 1/2 O_2 = H_2O$ is described by $\log K = 13152/T - 3.039$ (Chase, 1998) and is used to provide self-consistent solutions for $fH_2$ and $fH_2O$. An equation analogous to eq. (19) is also solved for $H_2$.

The results indicate that *i*) the total pressure of the sum of H-bearing species never exceeds ∼6 bar, *ii*) the dominant H-bearing species is $H_2$ below IW, and *iii*) the fraction of H degassed increases from 0.06 at ∆IW+3 to 0.15 at ∆IW-2, owing to the lower solubility of molecular hydrogen with respect to water.

As the partial pressures of gas species derived from silicates (*e.g.*, Mg(g), SiO(g)) are relatively low even at high temperatures ($10^{-3}$ bar at 2000 K, Charnoz et al., 2021), it is likely that the Moon never had a thick atmosphere (fewer than tens of bars; see also Needham and Kring, 2017 that estimate mare volcanism produces only 0.01 bar). Moreover, loss of H above a protolunar magma ocean would have been favoured under reducing conditions as not only does increasing $fH_2/fH_2O$ lower the mean molecular mass of the atmosphere from 17.5 at ∆IW+3 to 3.5 at ∆IW-2, but a factor ∼2 higher fraction of the budget of H resides in the atmosphere. Stated otherwise, should the Moon have had an initial $H_2O$ budget similar to that of Earth, and an oxidised magma ocean, then its water depletion could likely only have arisen owing to H loss in the protolunar disk, prior to its accretion. Hence, consideration of the $fO_2$ of the protolunar disk and above its magma ocean are paramount to assessing whether the scenario envisaged by Nakajima and Stevenson (2018) is applicable (above ∼ ∆IW+0.5), and therefore the ease with which the Moon dehydrated during its formation.

### 5.3. Prevalence of steam atmospheres around rocky bodies

A steam atmosphere is defined as one dominated by water vapour with respect to other gaseous species that may occur in planetary atmospheres at high temperatures (CO, $CO_2$, $H_2$). Such atmospheres are thought to have been relevant to the formation of the Earth in a magma ocean state (*e.g.*, Abe and Matsui, 1985; Fegley et al., 2016; Lebrun et al., 2013), to the terrestrial planets more generally (Lammer et al., 2020), as well as for exoplanets (Turbet et al., 2020). Indeed, the dissolution of water into silicate has been proposed to account, in part, for the 'radius gap' observed between super-Earths and sub-Neptunes (Zeng et al., 2021). However, whether steam atmospheres form is contingent upon the abundance and solubility of H with respect to other atmophile elements, of which the most abundant is C.

As demonstrated above, the fraction of H outgassed is not only dependent on intensive variables such as temperature and oxygen fugacity, but also extensive variables, namely, its abundance and planetary mass. To assess the prevalence and occurrence of steam atmospheres above magma oceans on Earth-like planets, a Monte Carlo simulation with 10,000 runs is performed across three variables that are uniformly distributed, the molar C/H ratio of bulk planet $\{C/H \in \mathbb{R} \mid 0.1 < C/H < 1\}$, the abundance of H in ocean masses, $N_{\text{Ocean}}$ $\{N_{Ocean} \in \mathbb{R} \mid 1 < N_{Ocean} < 10\}$ and the $fO_2$ at the magma ocean atmosphere surface $\{\Delta IW \in \mathbb{R} \mid -4 < \Delta IW < 4\}$. The partial pressures of the species CO, $CO_2$, $H_2$ and $H_2O$ are calculated in these model atmospheres at a constant temperature, 2173 K and planetary mass, that of the Earth (1 $M_E$, where 0.68 of this is its mantle) in a self-consistent manner according to mass balance constraints (eq. (19)) and homogeneous gas phase equilibria (see Bower et al., 2022, for a complete description).

At 2173 K, only the four major gas species in the system C-O-H shown in Fig. 9 are stable (*cf.* Sossi et al., 2020; Bower et al., 2022). A striking feature of the generated compositions is the mutual exclusivity of $CO_2$-$H_2$-rich atmospheres (Fig. 9). Within the explored parameter space, a maximum of roughly 75 bar $H_2$ and 75 bar $CO_2$ can coexist at intermediate $fO_2$ near the IW buffer (10 Ocean masses, C/H = 1), at which the major gas species is CO (∼200 to 400 bar). On average, CO is the most abundant species expected in the range of modelled atmospheres, despite the fact that bulk C/H ratios are ≤1. This is a product of its low solubility with respect to $H_2O$ and $CO_2$, and the fact that, at 2173 K, chemical equilibrium dictates that $CO/CO_2 = 1$ at IW+1.5 (with CO predominant below this value) as compared to IW for $H_2/H_2O = 1$. As a result, there is a wide parameter space of CO-$H_2$-rich atmospheres that occurs for all C/H ratios and ocean masses below the IW buffer. Considering that the formation of iron-rich cores on rocky planets in our Solar System sets the $fO_2$ to between ∆IW-1 and ∆IW-4 (*e.g.*, Frost et al., 2008), this condition may be commonplace among rocky planets. Nevertheless, the $fO_2$ set by core formation may not reflect that at the surface of the magma ocean, owing to the volume changes of redox equilibria, notably of iron (Hirschmann, 2012), such that higher $fO_2$ conditions at the interface are possible (Deng et al., 2020; Sossi et al., 2020).

Steam atmospheres *sensu stricto*, in which $fH_2O/(fH_2O + fH_2 + fCO + fCO_2) \geq 0.5$, do not form; the maximum value of the index achieved in the simulations is 0.34, and only 0.72% have values >0.25. Water, as the most soluble species amongst the four considered, reaches a maximum fugacity of only 37 bar (as compared with ∼700 bar for both $CO_2$ and $H_2$, and ∼500 bar for CO), meaning steam atmospheres are, on average, far more tenuous than the equivalent carbon-rich atmospheres produced at a given C/H ratio. Part of this behaviour derives from the dependence of dissolved $H_2O$ on $(fH_2O)^{0.5}$, one that holds until ∼1000 bar $fH_2O$ (*e.g.*, Berndt et al., 2002; Dixon et al., 1995). At pressures higher than this, significant quantities of $H_2O$ can more readily enter the atmosphere, as $K_{(6)}$ shifts to the left and $XH_2O$ becomes proportional to $fH_2O$. Care should therefore be taken in applying eq. (11) to magma oceans of planets with large $g$ and/or water budgets ($fH_2O > 1000$ bar), in which water dissolution would tend to a value of $\beta = 1$. Importantly, steam atmospheres for the scenarios considered are never composed entirely of steam; they contain significant quantities of $H_2$, $CO_2$ and CO. Indeed, the formation of steam atmospheres requires a bulk planet with >10 ocean masses, C/H ratios <0.1 and oxygen fugacities at the atmosphere-magma interface just above the IW buffer. This is because, although $fH_2O$ scales with $(fO_2)^{0.5}$, so too does the fugacity of the less soluble $CO_2$, meaning $fCO_2$ quickly surpasses $fH_2O$ above IW.

As a result, steam atmospheres are likely to be uncommon on Earth-sized rocky planets, lest they have molar C/H ratios significantly lower than that of Earth (∼1). In the solar neighbourhood, the C/H ratios of most G-type stars vary by ±0.5 *dex* units about the solar ratio (Hinkel et al., 2014). If this variability is indicative of that found in any potential orbiting rocky planets, then only a minority of stars, about 53 of 3819 G-type stars with a C/H ratio ∼10 times lower than that of the Sun, would be expected to host rocky planets that harbour steam atmospheres.

The foregoing assessment neglects other atmophile elements (notably S and N), which would further dilute the $H_2O$ fraction in the atmosphere, particularly at high $fO_2$ (>∆IW+2) at which $SO_2$ becomes a significant component (Gaillard et al., 2022). Moreover, only high temperature atmospheres (2173 K) in equilibrium with an Earth's mantle-sized magma ocean are considered. However, a variety of post-magma ocean processes have the potential to modify the constitution of the magma-generated atmosphere (Bower et al., 2022; Gaillard et al., 2022; Hier-Majumder and Hirschmann, 2017; Sossi et al., 2020). Notably, homogeneous gas phase equilibria, precipitation of graphite and other condensed phases, and their subsequent separation from the atmosphere during cooling can promote the production of $H_2O$ by increasing $fO_2$ and decreasing the C/H ratio, or its consumption through condensation (Sossi et al., 2020). Even prior to its outgassing, dynamical pro-cesses in the magma ocean, such as melt trapping (Hier-Majumder and Hirschmann, 2017) and dynamical lock-up (Bower et al., 2022) can prevent wholesale outgassing of $H_2O$, while fractional crystallisation prolongs its outgassing by maintaining contact between the melt and the atmosphere (Bower et al., 2022). Therefore, whether steam atmospheres around planets eventually arise during cooling and crystallisation depends upon *i*) efficient liberation of dissolved water and/or *ii*) chemical equilibration of the atmosphere and its components upon cooling.

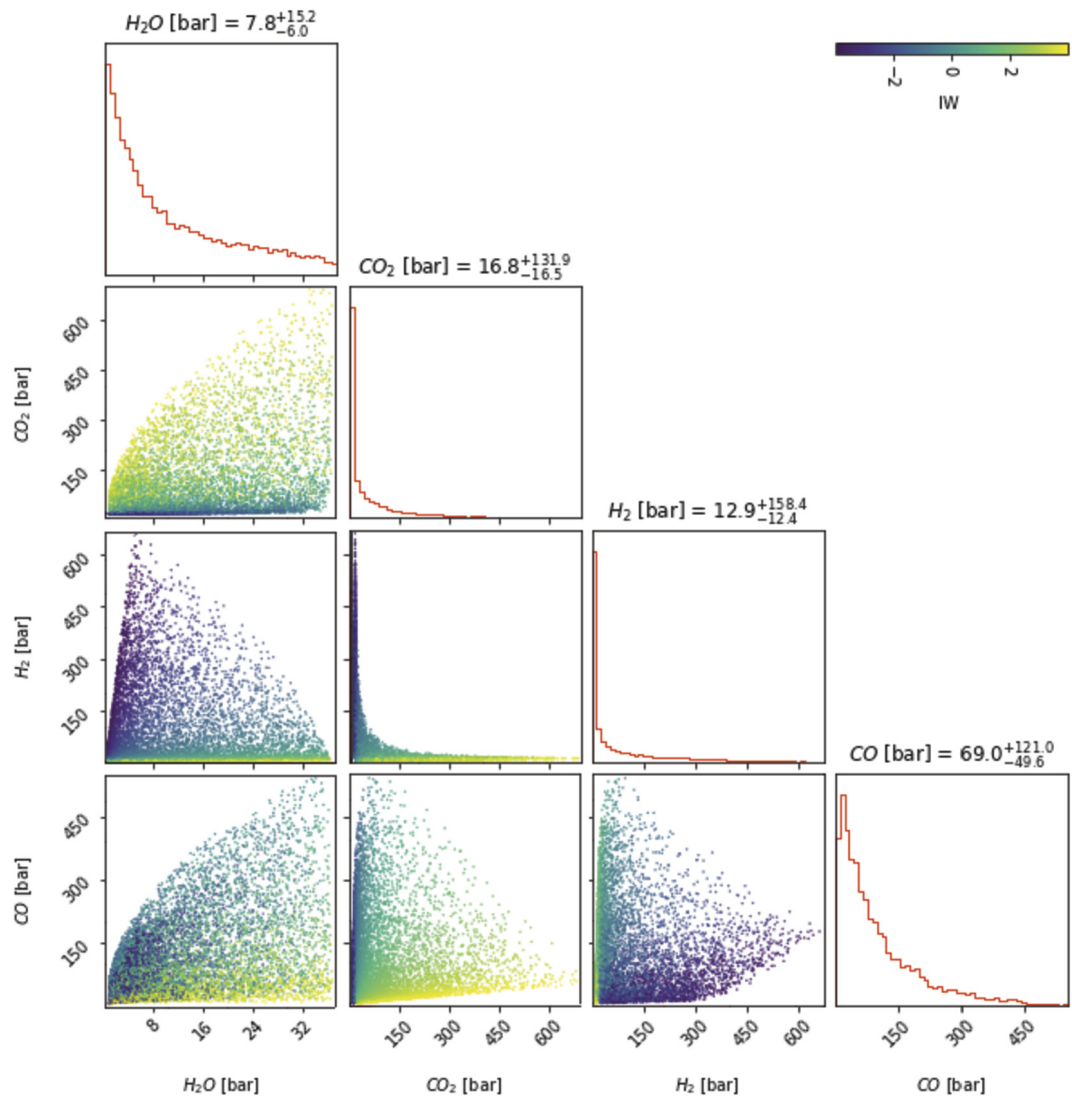

**Fig. 9.** Corner plot of the partial pressures of the four gas species present in the atmospheres of planets with 1 Earth mass ($M_E$) and a fully-molten mantle of 0.68 $M_E$ at constant temperature (2173 K), generated according to a Monte-Carlo simulation by variation of three parameters – C/H ratio, ocean masses of H, and oxygen fugacity (see text for details). The colours of the points are coded according to the $fO_2$ value, reported with respect to the iron-wüstite buffer ($\Delta$IW). Equivalent plots in which the points are colour-coded by C/H ratio and ocean masses are shown in the Supplementary Information. Histograms are constructed with 50 bins, and the title shows the 0.5 quantile of the data, with lower and upper bounds that pertain to the 0.16 and 0.84 quantiles, respectively. An $\alpha_{H_2O}$ of 524 ppmw/bar$^{0.5}$ is used (use of $\alpha_{H_2O}$ = 647 ppmw/bar$^{0.5}$ would result in even lower pressures of $H_2O$).

## 6. Conclusions

The generation and prevalence of steam atmospheres around rocky planets is an oft-invoked, yet poorly understood phenomenon. This is largely owing to the scarcity of solubility data for compositions, and at pressures and temperatures, relevant to the formation of atmospheres from molten planetary mantles. We demonstrate, through experimental determination of $H_2O$ contents in peridotite liquids at 2173 K using Fourier-Transform InfraRed spectroscopy, that the dissolved content of $H_2O$ is proportional to the square-root of the fugacity of $H_2O$ with which it is in equilibrium. This dependence confirms its incorporation as hydroxyl groups (OH) in silicate melts at the low total $fH_2O$ (up to 0.027 bar) of the experiments. Moreover, $fH_2$ promotes dissolution of OH at low $fO_2$ below the iron-wüstite buffer. At a given $fH_2O$, temperature exerts the primary control on water solubility in silicate liquids, and the water contents of peridotite melts at 2173 K are 10 - 25% lower than for basaltic melts at 1623 K at equivalent $fH_2O$, offsetting the effect of lower melt polymerisation, which is expected to increase $H_2O$ solubility. Peridotite liquids are able to dissolve $H_2(g)$ as OH in their structure, with a coefficient 1/3 that for dissolution of $H_2O(g)$, a property linked to their highly depolymerised nature.

We demonstrate that a greater fraction of the total water budget of a given planetary body is outgassed for smaller with respect to larger bodies, owing to the square-root dependence of $fH_2O$ on dissolved $H_2O$. This property, together with the reducing $fO_2$ of basaltic achondrites compared to terrestrial basalts, provides an explanation for the systematically low $H_2O$ contents of nominally anhydrous minerals in achondritic meteorites derived from the crusts and mantles of small telluric bodies with respect to their terrestrial counterparts. Assuming an initially Earth-like H budget, one such small body, the Moon, likely only ever harboured a tenuous atmosphere, with less than several bars of H as $H_2$ (∼6 bar, below IW) or $H_2O$ (∼2 bar, above IW). The $fO_2$ that prevailed in the protolunar atmosphere affects its mean molar mass, and is therefore key to determining whether H is retained as $H_2O$ or more readily escapes as $H_2$. A survey of the possible chemical variability among rocky (exo)planets shows that the $fO_2$ at the atmosphere-magma interface, together with the C/H ratio and number of oceans in the bulk planet, control the likelihood of steam atmosphere formation at magmatic temperatures. We show, through Monte-Carlo simulation, that the high solubility of $H_2O$ with respect to other gases in the C-O-H system limits the extent of its atmospheric outgassing. Therefore, steam atmospheres are rare outcomes of magma ocean-derived secondary atmospheres, and require low C/H ratios, and high H budgets and oxygen fugacities to prevail.

**CRediT authorship contribution statement**

**Paolo A. Sossi:** Conceptualization, Formal analysis, Investigation, Writing – original draft, Writing – review & editing. **Peter M.E. Tollan:** Formal analysis, Investigation, Writing – review & editing. **James Badro:** Investigation, Resources, Writing – review & editing. **Dan J. Bower:** Formal analysis, Writing – review & editing.

**Declaration of competing interest**

The authors declare that they have no known competing financial interests or personal relationships that could have appeared to influence the work reported in this paper.

**Data availability**

Raw data is available as csv files in the supplementary information.

**Acknowledgements**

We are grateful to Jörg Hermann and Mona Lüder for providing access and support at the FTIR facility at the Universität Bern. We gratefully acknowledge Dmitry Bondar, Tony Withers and Tomoo Katsura in sharing their preliminary data for the molar absorption coefficient of the 3550 $cm^{-1}$ band in hydrous peridotite glasses with us. The constructive, impartial and insightful comments of three anonymous reviewers contributed to improvements to many aspects of this work. We greatly appreciate the efficient editorial handling of Rajdeep Dasgupta. P.A.S. thanks the Swiss National Science Foundation (SNSF) for support via Ambizione Fellowship #180025 and the Swiss State Secretariat for Education, Research and Innovation (SERI) under contract number MB22.00033, a SERI-funded ERC Starting Grant '2ATMO'. D.J.B. thanks the SNSF for support via Ambizione Fellowship #173992.

**Appendix A**

We solve the interior static structure equations (*e.g.*, Valencia et al., 2007) to determine the structure and bulk properties of differentiated planetesimals and planets with radii ranging from around 10 km to 7700 km. To this end, we first prescribe the bulk planetary mass and a fixed core mass fraction of 30% across all radii, where the density of the iron core is assumed to be constant (pressure-independent) at 10000 $kg/m^3$. This approach enables determination of the radius of the core-mantle boundary and the mass of the mantle. The remaining step is to determine the radius of the planet, $r_0$, given the mass of the mantle and core, the radius of the core-mantle boundary, and an equation of state for the silicate mantle. To capture the density change ($d\rho$) of peridotite with radius ($dr$) we assume adiabatic self-compression (the Adams-Williamson equation),

$$\frac{d\rho}{dr} = -\frac{\rho(r)^2 g(r)}{K}, \quad (A.1)$$

where $\rho$ is the density, $g$ is the acceleration due to gravity, and $K$ is the adiabatic bulk modulus of the mantle. Here we employ a value of $K = 264$ GPa determined for bridgmanite (Jackson, 1998) to approximate that of peridotite. The remaining static structure equations are:

$$\frac{dg}{dr} = 4\pi G \rho(r) - \frac{2Gm(r)}{r^3}, \quad (A.2)$$

$$\frac{dm}{dr} = 4\pi r^2 \rho(r), \quad (A.3)$$

$$\frac{dP}{dr} = -\rho(r) g(r). \quad (A.4)$$

In the above equations, $G$ is the universal gravitational constant, $m$ is the mass contained within $r$ and $P$ is the pressure. To solve the system of equations we prescribe boundary conditions at the planetary radius ($r_0$) and integrate to the core-mantle boundary:

$$g(r_0) = Gm_0/r_0^2, \quad m(r_0) = m_0, \quad P(r_0) = 0, \quad \rho(r_0) = \rho_0 \quad (A.5)$$

We adopt a surface density $\rho_0$ of 3300 $kg/m^3$ for peridotite (Lee, 2003). Using Newton's method we then solve the structure equations for the planetary radius ($r_0$) that results in a silicate mantle that constitutes 70% of the prescribed bulk planetary mass. For the size range of planetary bodies we consider, this provides a self-consistent mantle & bulk planetary mass, planetary radius, and surface gravity to use in eqns. (17) and (18).

**Appendix B. Supplementary material**

Supplementary material related to this article can be found online at https://doi.org/10.1016/j.epsl.2022.117894.